\documentclass[pra,aps]{revtex4}
\usepackage{graphicx,amsfonts,bm,amsmath}
\newcommand{\kx}{|\mathrm{X}\rangle}
\newcommand{\ky}{|\mathrm{Y}\rangle}
\newcommand{\kup}{|+\rangle}
\newcommand{\kdn}{|-\rangle}
\newcommand{\kbi}{|\mathrm{XX}\rangle}
\newcommand{\kg}{|\mathrm{g}\rangle}

\newcommand{\bx}{\langle\mathrm{X}|}

\newcommand{\pbi}{|\mathrm{XX}\rangle\!\langle\mathrm{XX}|}
\newcommand{\pg}{|\mathrm{g}\rangle\!\langle \mathrm{g}|}

\begin{document}
\title{Two-photon coherent polarization flipping of confined excitons}
\author{P. Machnikowski}
 \affiliation{Institute of Physics, Wroc{\l}aw University of
Technology, 50-370 Wroc{\l}aw, Poland}

\begin{abstract}
\begin{center}
\parbox{125mm}{
A two-photon process leading to coherent transitions between the two
circularly polarized exciton states in a quantum dot is studied. It is
shown that optical flipping of the exciton polarization is possible
with picosecond laser pulses. The process is closely related to
two-photon Rabi oscillations of a biexciton but it is much more stable
against shifts of the laser frequency.
}
\end{center}
\end{abstract}


\maketitle

Recent years have witnessed enormous progress in the optical control
of charge states in semiconductor quantum dots (QDs). For instance,
coherent optical transitions (Rabi oscillations) 
between the ground state and a
single-exciton state of a single QD have been demonstrated
\cite{zrenner02}. This and other experiments show that quantum optical
schemes can be implemented on these artificial systems.
Another experimental achievement in this field is the observation
of two-photon Rabi oscillations between the ground and biexciton
states of a single self-assembled QD \cite{stufler06}. This experiment
has shown that coherent control of two-photon transitions in QDs far
beyond the perturbative regime is feasible. 

On the other hand, the dynamically developing field of spintronics and
spin-based implementations of classical and quantum information
processing has motivated the search for fast, optical methods of
control for confined electron spins
\cite{troiani03,chen04,nazir04,lovett05}.  
Experiments in this area
\cite{dutt05,greilich06c}, even though still limited to QD ensembles,
seem very promising. 

This paper extends the theory of two-photon optical control of a
biexciton system \cite{stufler06}
and shows that two-photon transitions may be used to coherently
rotate the spin orientation (polarization) of a single exciton
confined in a QD. 

\begin{figure}[tb]
\begin{center}
\unitlength 1mm
\begin{picture}(110,40)(0,5)
\put(0,0){\resizebox{36mm}{!}{\includegraphics{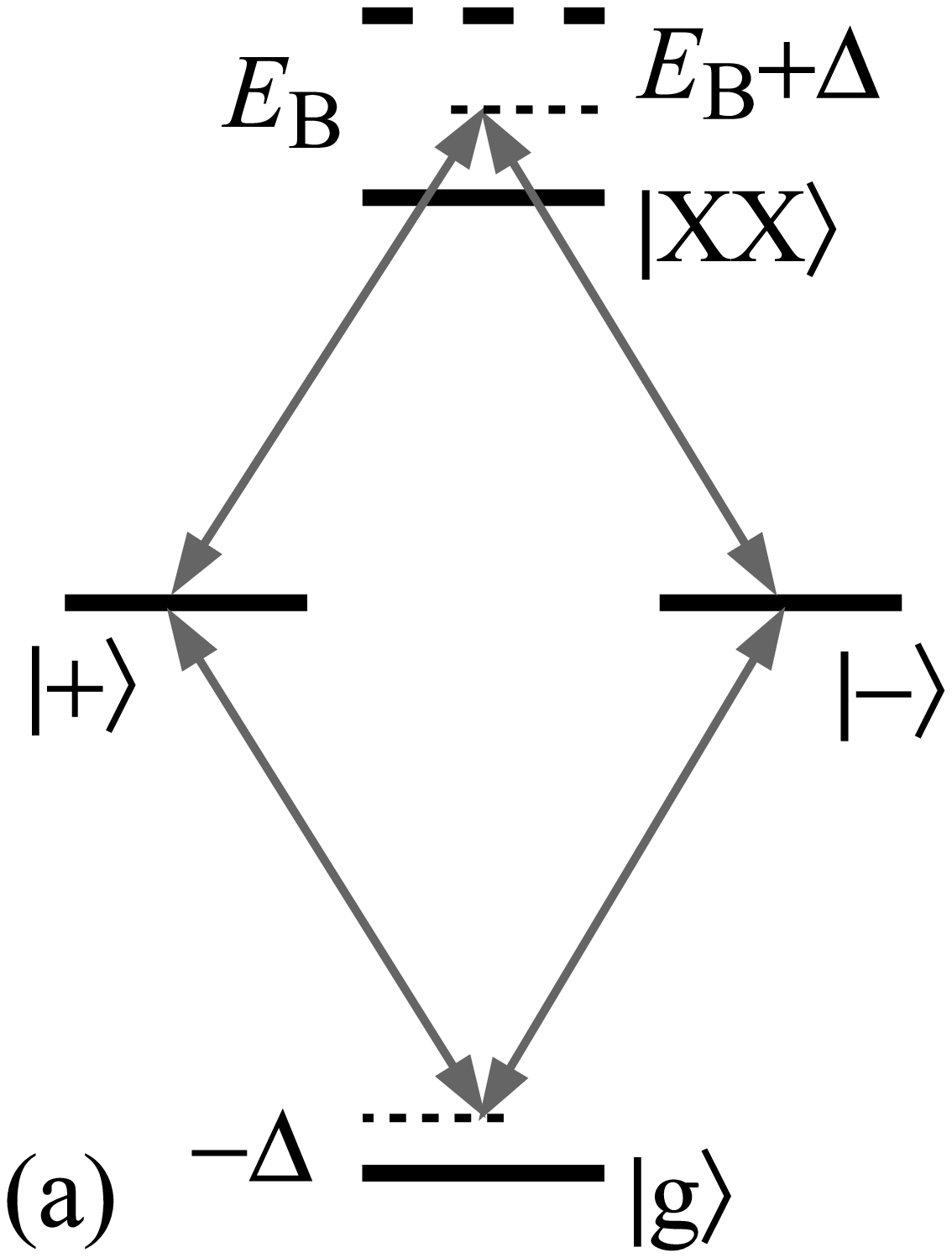}}}
\put(55,0){\resizebox{55mm}{!}{\includegraphics{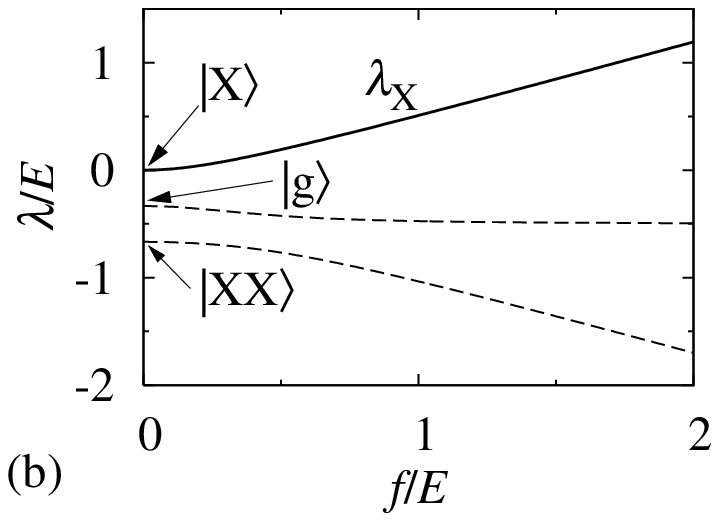}}}
\end{picture}
\end{center}
\caption{\label{fig:1}(a) The energy level structure of a confined
biexciton and the schematic presentation of the optical couplings
between the states. (b) The spectral branches representing the
instantaneous eigenstates of the Hamiltonian as functions of the
pulse amplitude $f$ (the adiabatic parameter). Here
$\Delta=-E_{\mathrm{B}}/3$, $E_{\mathrm{B}}>0$.} 
\end{figure}

Let us consider a single QD coupled to a linearly
polarized laser beam. According to the selection rules, such a beam
couples both the exciton states to the ground and biexciton
states. We assume that the frequency of the beam is detuned from both
these transitions, as shown in Fig.~\ref{fig:1}a. For pulses of
picosecond durations the excited states of confined carriers are
irrelevant and may be disregarded. We will also neglect the exchange
interaction that couples the two circularly polarized exciton states
and turns them into a weakly split linearly polarized doublet
\cite{ivchenko97}. The energy of this coupling is usually below
100~$\mu$eV and does not affect the system dynamics on the time scales
relevant for the present discussion. 

Thus, in the rotating wave approximation the system is described by
the Hamiltonian 
\begin{equation}
\label{ham}
H=-(E_{\mathrm{B}}+\Delta)\pbi+\Delta\pg+
\frac{1}{2}f(t)\left[ (\kg+\kbi)\bx +\mathrm{H.c.}\right],
\end{equation}
where $\kg,\kbi$ denote the ground and biexciton states,
$\kx=(\kup+\kdn)/\sqrt{2}$ is the bright
(``$x$-polarized'') superposition of the
two single-exciton states $\kup,\kdn$ with definite circular
polarizations, $E_{\mathrm{B}}$ is the biexciton shift (assumed
positive in the binding case, as in Fig.\ref{fig:1}), $\Delta=\omega-E$ is the
detuning between the laser frequency $\omega$ and the single-exciton
transition energy $E$, and $f(t)$ is the envelope of the laser pulse
amplitude. The other, dark (``$y$-polarized'') superposition of the
exciton states, $\ky=(\kup-\kdn)/\sqrt{2}$, remains decoupled. 

We will assume that the system is initially prepared in one of the
single-exciton states, say $\kup$. This state can be represented as 
$\kup=(\kx+\ky)/\sqrt{2}$. The state $\ky$ is decoupled from the laser
beam and does not evolve. The evolution of the state $\kx$ under
appropriate conditions may be found using the adiabatic theorem
\cite{messiah66,stufler06}.  
To this end, let us consider the diagram of instantaneous eigenstates
of the relevant three-level Hamiltonian (excluding the decoupled state
$\kdn$) as a function of the pulse amplitude $f$, shown in
Fig.\ref{fig:1}b. As can be seen, in general the spectral branch containing the
state $\kx$ remains non-degenerate and separated from the other states
by at least $\min (|\Delta|,|\Delta+E_{\mathrm{B}}|)$. Here, we will
only consider the situation when the detunings of the laser frequency from
both transition energies, $|\Delta|$ and $|\Delta+E_{\mathrm{B}}|$ are
much larger that both the maximum pulse amplitude and the pulse
spectral width. In such case, the state $\kup$ undergoes an adiabatic
evolution, with $f(t)$ playing the role of a slowly varying
parameter. After the pulse has been switched off, the initial state is
restored with the additional dynamical phase 
\begin{displaymath}
\alpha=-\frac{1}{\hbar}\int_{-\infty}^{\infty}dt \lambda_{\mathrm{X}}(t),
\end{displaymath}
where $\lambda_{\mathrm{X}}(t)$ is the instantaneous eigenstate of the
Hamiltonian 
(\ref{ham}) corresponding to the state $\kx$ (see Fig.\ref{fig:1}b).
As a result, the state $\kup$ undergoes the transformation
\begin{displaymath}
\kup=\frac{\kx+\ky}{\sqrt{2}}\longrightarrow 
\frac{e^{i\alpha}\kx+\ky}{\sqrt{2}}
=e^{i\alpha/2}
\left( \cos\frac{\alpha}{2}\kup+i\sin\frac{\alpha}{2}\kdn \right).
\end{displaymath}
The phase $\alpha$ may be arbitrarily large. Moreover, for a fixed
pulse shape it is a monotonous function of the pulse intensity. Thus,
by varying the pulse amplitude, 
the exciton can be coherently rotated between the two polarization
states $\kup$ and $\kdn$, much like in the usual pulse-area dependent Rabi
oscillation between the ground and single exciton states, induced by a
resonant circularly polarized beam \cite{zrenner02}.

The instantaneous eigenvalue $\lambda_{\mathrm{X}}(t)$ results from
diagonalizing a 
three-level Hamiltonian which can always be done
analytically. However, the resulting formula is reasonably simple only
in the symmetrically detuned case $\Delta=-E_{\mathrm{B}}/2$, when one
finds
\begin{equation}\label{lambda}
\lambda_{\mathrm{X}}(t)
=E_{\mathrm{B}}\frac{\sqrt{1+8[f(t)/E_{\mathrm{B}}]^{2}}-1}{4}.
\end{equation}
In the general case, it seems more reasonable to calculate the system
evolution numerically.  

\begin{figure}[tb]
\begin{center}
\unitlength 1mm
\begin{picture}(110,50)(0,5)
\put(0,0){\resizebox{110mm}{!}{\includegraphics{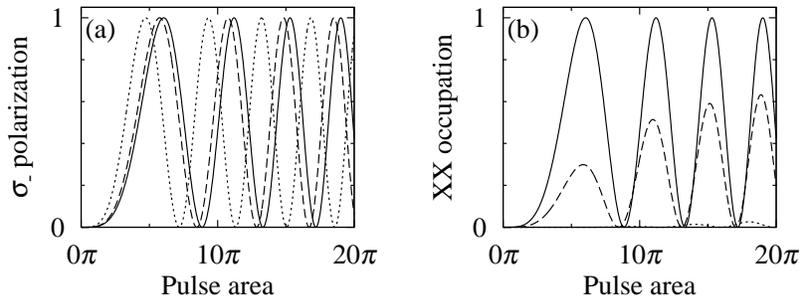}}}
\end{picture}
\end{center}
\caption{\label{fig:2}(a) Two-photon polarization flip: 
the occupation of the state $\kdn$
($\sigma_{-}$-polarized exciton) as a function of the pulse area, for
$\tau_{0}=5$~ps, $E_{\mathrm{B}}=4$~meV, and $\Delta=E_{\mathrm{B}}/2$
(solid), $E_{\mathrm{B}}/3$ (dashed), and $E_{\mathrm{B}}/6$
(dotted). (b) Two-photon Rabi oscillations \cite{stufler06}: the
occupation of the 
biexciton state as a function of the pulse area, for
$\tau_{0}=5$~ps, $E_{\mathrm{B}}=4$~meV, and $\Delta=E_{\mathrm{B}}/2=2$~meV
(solid), $\Delta=1.9$~meV (dashed), and $\Delta=1.7$~meV
(dotted).}
\end{figure}

The final occupation of the $\kdn$ state as a function of the pulse area
\begin{displaymath}
\theta=\int_{-\infty}^{\infty}dt f(t)
\end{displaymath}
for various detuning conditions is shown in Fig.~\ref{fig:2}(a). 
In all the cases,
the exciton undergoes oscillations between the two polarization
states. In spite of some qualitative similarity to the Rabi
oscillations in a two-level system 
(described by the universal function $\sin^{2}\theta/2$) 
one can clearly see essential differences. 
The oscillations are not strictly periodic,
especially for weak pulses, where the transition probability develops
very slowly. In fact, for weak pulses the occupation of the flipped
state grows as $\theta^{4}\sim I^{2}$, where $I$ is the pulse
intensity, as expected for a two-photon process (see
Ref.~\cite{stufler06}). 
Moreover, 
it is clear from Eq.~(\ref{lambda}) that the rotation angle is a nonlinear
functional of the pulse envelope. Hence, contrary to the usual Rabi
oscillations, no universal area theorem exists for the final occupations.

Let us note that the symmetrical detuning condition is equivalent to
the two-photon resonance between the ground and biexciton
states. Thus, in this case the two-photon polarization flip results
from the same evolution as the two-photon Rabi oscillations 
of the biexciton \cite{stufler06} (only the initial state is
different). The very good agreement between the description based on the
adiabatic theorem and the experimental results \cite{stufler06} shows
that the theory captures the essentials of the quantum evolution under actual
laboratory conditions (except for dephasing effects that are obviously
beyond the scope of the present theory). 

There is, however, a fundamental difference between the two-photon
biexciton oscillations and the polarization flip. If the two-photon
resonance condition $\Delta=E/2$ is not satisfied, the transition from
the ground to biexciton state is forbidden and the corresponding Rabi
oscillations are precluded. On the contrary, the polarization flip is
always resonant. For comparison of the two processes the results for
two-photon Rabi oscillations are plotted in Fig.~\ref{fig:2}(b). While
the curves describing the two processes at the two-photon resonance
(symmetric detuning) are the same, they become very different out of
resonance. This additional freedom of control may be useful for
optimizing experimental parameters against phonon-induced dephasing
\cite{krugel05,roszak05b}.

The two-photon polarization flip described here 
is expected to take place under the same conditions as
the two-photon Rabi oscillations that have already been
demonstrated in an experiment \cite{stufler06}. Moreover,
the polarization of an exciton is relatively easily
measurable. Therefore, it should be possible to observe the two-photon
polarization flip using currently available experimental
techniques. This would provide an additional degree of control over the
biexciton system which can be treated as a two-bit quantum
register if the logical values of 0 and 1 are associated with the
absence and presence, respectively, of an exciton with one of the two
circular polarizations. Like two-photon Rabi oscillations, the
two-photon 
polarization flip is an entangling operation, with maximally entangled
states of the two excitons, 
of the form $(|01\rangle\pm i|10\rangle)/\sqrt{2}$, 
reached for $\alpha=(n+1/2)\pi$.

%\newpage

%\bibliographystyle{prsty}
%\bibliography{abbr,quantum}

%\newpage

%\section*{Figure captions}
%\setcounter{figure}{0}

\end{document}